\documentclass[twocolumn]{article} 
\usepackage{amsmath}
\usepackage{graphicx}
\usepackage{color}
\usepackage{epstopdf}
\usepackage{url}
\usepackage[affil-it]{authblk}

\newcommand{\eqnRef}[1]{eq.~\ref{eqn:#1}}

\newcommand{\figRef}[1]{fig.~\ref{fig:#1}} % Copying Terry's formatting
\newcommand{\figRefs}[2]{figs.~\ref{fig:#1} and~\ref{fig:#2}} % Two figures

\newcommand\vopar        {\boldsymbol{M}}
\newcommand\opar         {M}
\newcommand\IOP          {\mathcal I}
\newcommand\Ipw          {\IOP_{pw}}
\newcommand\IpwLT        {\IOP^{LT}_{pw}}
\newcommand\TOP          {\mathcal T}
\newcommand\Tgl          {\TOP_{gl}}
\newcommand\TglLT        {\TOP^{LT}_{gl}}
\newcommand\TglTD        {\TOP^{2D}_{gl}}
\renewcommand\phi        {\varphi}
\newcommand{\entrop}      {{\mathbf H}}
\newcommand\bTheta       {\boldsymbol\Theta}

\providecommand{\e}[1]{\ensuremath{\times 10^{#1}}}
\newcommand{\ie}[0]      {\textit{i.e.}}
\providecommand{\etal}[0]    {\textit{et al.}}

\newcommand{\nicesim}{{\raise.17ex\hbox{$\scriptstyle\mathtt{\sim}$}}}

\begin{document}
\title{Information flow in finite flocks with topological interactions}
% \shorttitle{Title} %Insert here a short version of the title if it exceeds 70 characters

\author{J. Brown \thanks{Email: \texttt{jbrown@csu.edu.au}; Corresponding author}} \affil{School of Computing \& Mathematics, Charles Sturt University, Bathurst, NSW, Australia}
\author{T. Bossomaier} \affil{Centre for Research in Complex Systems, Charles Sturt University, Bathurst, NSW, Australia}
\author{L. Barnett} \affil{Sackler Centre for Consciousness Science, Department of Informatics, University of Sussex, Brighton, U.K.}

\maketitle

\begin{abstract}
We simulate the Vicsek model utilising topological neighbour interactions and estimate information theoretic quantities as a function of noise, the variability in the extent to which each animal aligns with its neighbours, and the flock direction. We show that these quantities, mutual information and global transfer entropy, are in fact dependent on observation time, and in comparison to the canonical Vicsek model which utilises range-based interactions, the topological variant converges to the long-term limiting behaviour with smaller observation windows. Finally, we show that in contrast to the metric model, which exhibits maximal information flow for the ordered regime, the topological model maintains this maximal information flow beyond the phase transition and into the disordered regime.
\end{abstract}

\section{Introduction}
\label{sec:intro}

The scale and grace of bird flocks are nature's most impressive displays, arising from seemingly chaotic flight paths of individual birds. Flocking behaviour is not restricted to just birds, either, with many other species displaying similar movements, from schools of fish~\cite{Shaw78} to colonies of bacteria~\cite{Keller71}. The phenomenon of collective motion, and its constituent components, are subject to much research~\cite{Vicsek12}.

While new stereographic camera techniques and equipment allow research into large real-world flocks, much of the literature continues to use abstract models, such as the Standard Vicsek Model~(SVM)~\cite{Vicsek95} or the Inertial Spin Model~\cite{Cavagna15}. These models approximate real-world systems as point particles whose decision making processes are encapsulated by local neighbour interaction rules and errors (\emph{noise}). The SVM is perhaps the most minimal such model in that it contains only a single rule: \emph{assume the average direction of the local neighbourhood of radius $r$ with some random perturbation added}~\cite{Vicsek95}, where the magnitude of the noise determines whether the flock is coherent or not. 

An implication of local interaction rules is that observing one flock member can reveal information about its nearby flock mates---for instance, we can better guess, or be more certain of, neighbouring headings. Information Theory allows this reduction in uncertainty to be quantified. Mutual Information~(MI)~\cite{Shannon48} defines \emph{information sharing}, the symmetric and instantaneous reduction in uncertainty about the heading of one flock member when any other member is observed. \emph{Information flow} instead quantifies the temporal reduction in uncertainty between flock members, that is, how knowledge of the current heading affects estimation of a future heading, which we measure here with Global Transfer Entropy (GTE)~\cite{Barnett13}.

Information sharing among flock members is crucial to flock formation and stability, yet in the SVM it behaves unexpectedly in these finite flocks~\cite{Barnett18,gte-paper}: MI is shown to diverge as flocks become increasingly ordered, while information flow converges to a finite, non-zero value, where both are expected to vanish in a highly ordered flock.

The unexpected behaviour in the SVM is due to continuous symmetry in the system. Thermal fluctuations allow the flock orientation of a highly ordered flock to proceed on a random walk without affecting overall order. When observed over small periods of time, an ordered flock is confined to a small region of phase space---that is, the system is restricted to only a small number of all possible states---producing more expected results---namely, MI and GTE tending to $0$ as flocks become more ordered, as seen in~\cite{Wicks07} for MI. However as observation time is increased, flock orientation gradually drifts on a random walk, eventually becoming uniform over $2\pi$ in the long-term (infinite) limit, exploring the entire phase space along the way. When the system is confined in phase space, it is said to have broken \emph{ergodicity}; the assumption that behaviour averaged over time is the same when averaged over phase space. Thus over short observation periods, the SVM breaks ergodicity, while in the long-term observation limit, it is restored, leading to continuously-broken ergodicity~\cite{Mauro07} and the phenomena of diverging MI and non-zero GTE.

The SVM flock members interact with neighbours within a fixed radius, \ie, metric interactions. However, it is becoming increasingly apparent that this is not the case for many real-world flocks. Ballerini~\etal~\cite{Ballerini08} show via detailed 3D recordings of starlings that birds in real-world flocks instead interact in a topological manner. That is, a bird will interact with its closest six to seven neighbours, regardless of distance. Similar work has shown that 3D schools of fish, and (effectively) 2D herds of sheep and deer also utilise topological interactions~\cite{Gautrais12,Kattas13}. \cite{Niizato11} further show that in ``imperfect'' biological organisms there might not be a strict delineation between the two, as both interaction methods can be seen as ``uncertain'' approximations of each other.

Here we apply the information theoretic metrics, MI and GTE, to a topological variant of the Vicsek Model (TVM) in which flock members take the average direction of their $k$ closest neighbours rather than those within $r$ units.

By noting a fundamental instability in the TVM over long time scales~\cite{Brown17a}~we show here that the information theoretic behaviours of TVM are not only comparable to those of its metric counterpart, but in fact converge to the long-term limiting behaviours with shorter observation windows.

\section{The topological Vicsek model}
\label{sec:vicsek}

The two dimensional TVM comprises a set of $N$ point particles (labelled $i=1,\ldots ,N$) moving on a plane of linear extent, $L$, with periodic boundary conditions. Each particle moves with constant speed, $v$, and interacts only with the closest $k$ neighbours. Positions, $\vec{x}_i(t)$, and headings, $\theta_i(t)$, are updated synchronously at discrete time intervals $\Delta t= 1$ according to

\begin{align}
  \vec{x}_i(t + \Delta t) &= \vec{x}_i(t) + \vec{v}_i(t)\Delta t\,, \label{eqn:OVAposUpdate} \\
  \theta_i(t+\Delta t) &= \phi_i(t) + \omega_i(t)\,, \label{eqn:OVAvelUpdate}
\end{align}
respectively, where $\vec{v}_i(t)$ is constructed from $\theta_i(t)$ and the constant $v$, $\phi_i(t)$ is the average (consensus) heading of the $k$ closest particles to $i$ (including itself), and $\omega_i(t)$ is the white noise uniform on the interval $[-\eta/2,\eta/2]$ with intensity $\eta \in [0, 2\pi]$ with the system exhibiting order at low noise, disorder at high noise and a transition between the two phase at some intermediate noise magnitude $\eta_c$. Note that the TVM is a flocking \emph{system} not a single flock, and that at the phase transition many flocks will exist, continually forming and breaking apart. Particle density $\rho=N/L^2=0.25$ is fixed throughout with $k=6$ in most experiments.

\cite{Barnett18,gte-paper} consider the TVM as a steady-state \emph{statistical ensemble} of finite size containing $N$ particles with control parameter $\eta$. Capitals indicate quantities sampled from the ensemble; particularly $\Theta_i$ is the ensemble sample of the heading of the $i$th particle. The 2D mean particle velocity vector $\vopar$ gives the \emph{order parameter}, with magnitude $\opar \in [0,1]$ and heading $\Phi \in (0,2\pi]$. $M = 1$ indicates complete order, with all particle headings aligned, while the disordered case of $\eta=2\pi$ gives $\opar \to 0$. The ensemble variance
\begin{equation}
 \chi = \langle \opar^2\rangle-\langle \opar\rangle^2
\label{eqn:suscep}
\end{equation}
defines the \emph{susceptibility}; a peak in $\chi$ as a function of $\eta$ is taken to locate an (approximate) phase transition~\cite{baglietto08:finite}.

When estimating ensemble statistics from steady-state dynamics one usually invokes \emph{ergodicity}: statistics are collated over an observation time window of length $T$ under the assumption that as $T\to\infty$ the statistic converges to its ensemble average value. This also assumes that $T$ dominates internal dynamical time scales in the system. This likely does not hold in the SVM~\cite{Barnett18}, and may not hold in the TVM either. Over short observation windows, the Vicsek model breaks symmetry as all headings collapse to some group consensus, however as observation time is increased, this heading performs a random walk, exploring progressively larger volumes of phase space. This in effect restores ergodicity over very large $T$, thus exhibiting behaviour akin to ``continuously-broken ergodicity''~\cite{Mauro07}.

As such, we consider two regimes when collecting ensemble statistics---\emph{short-term} and \emph{long-term}. In the former, we collate statistics separately---without any ergodic assumptions, \ie, an ensemble size of one---while in the long-term regime, we employ dimensional reductions developed in~\cite{Barnett18,gte-paper} (Discussed below) to explore the limiting behaviour of the TVM.

\section{Information Theoretic Quantities}
\label{sec:infoTheory}
Mutual Information gives the shared information between pairs of particles and is defined as the ensemble statistic
\begin{equation}
  \Ipw \equiv \IOP(\Theta_I : \Theta_J) = \entrop(\Theta_I) + \entrop(\Theta_J) - \entrop(\Theta_I,\Theta_J) \,,\label{eqn:Ipw}
\end{equation}
where $\entrop=\int p(x)\log p(x)dx$ denotes differential entropy and $(I,J)$ is uniform on the set of unique neighbour index-pairs. While differential entropy is somewhat ill-defined in that it can be negative, MI is a strictly non-negative quantity.

The ergodic assumption in the long-term limit allows for exploitation of rotational symmetry in the system~\cite{Barnett18} providing a novel dimensional reduction---involving just heading differences of neighbour pairs---for the long-term MI
\begin{equation}
\IpwLT = \log 2\pi - \entrop(\Theta_I - \Theta_J)\,.
\label{eqn:mi1D}
\end{equation}
At high noise, $\Theta_I - \Theta_J$ becomes uniform over $2\pi$ and thus $\IpwLT$ vanishes, as expected. At low noise however, particles align ever closer, leading to a sharp peak in $\Theta_I - \Theta_J$. Since this is differential entropy, $\entrop(\Theta_I - \Theta_J)$ will diverge to $-\infty$ as $\eta \to 0$. See~\cite{Barnett18} for a complete discussion of this reduction.

Information flow on the other hand is given by the Transfer Entropy~\cite{Schreiber00} of two random processes, $X$ and $Y$:
\begin{equation}
\TOP_{Y \to X} = \entrop(X' \,|\, X) - \entrop(X' \,|\, Y)\,,
\label{eqn:te}
\end{equation}
where information flows from $Y$ to $X$, $X'$ is the future state of $X$ and $\entrop(\cdot \,|\, \cdot)$ denotes conditional entropy and the GTE is defined as the ensemble statistic:
\begin{equation}
  \Tgl \equiv \TOP_{\bTheta \to \Theta_I} =  \entrop(\Theta'_I \,|\, \Theta_I) - \entrop(\Theta'_I \,|\, \bTheta) \label{eqn:Tgl}
\end{equation}
where $\bTheta=(\Theta_1,\ldots,\Theta_N)$ is the vector of all $N$ particle headings, shown in~\cite{gte-paper} to reduce to
\begin{equation}
  \TglTD  =  \entrop(\Theta'_I \,|\, \Theta_I) - \entrop(\Omega)\,. \label{eqn:Tgl2D}
\end{equation}

The above rotational symmetry assumption and reduction can be applied to the GTE in the long-term limit as well, providing a similar form to \eqnRef{mi1D}:
\begin{equation}
\TglLT = \entrop([\Theta'_I-\Theta_I]) - \entrop(\Omega)\,,
\label{eqn:gte1D}
\end{equation}
which relies purely on the change in a particle's heading over time, where $[\cdot]$ denotes the internal angle. As with $\IpwLT$, $\TglLT$ tends to $0$ in the high noise case. However here, both terms diverge as $\eta \to 0$, which results in a convergence to a finite, non-zero information flow, shown to be $\nicesim 0.72$ bits in simulation and $\nicesim 0.75$ bits when considering $[\Theta'_I-\Theta_I]$ as an approximately Gaussian variable with variance twice that of the noise~\cite{gte-paper}.

The above derivations were developed with the SVM in mind, however the assumptions made are independent of the particle neighbourhood method, and thus apply directly to the topological case, with results below.

Finally, the long-term limit quantities were in stark contrast to the short-term quantities, in which $\Ipw$ and $\Tgl$ did in-fact peak at or before the phase transition, as observed in other studies and systems. However, neither $\Ipw$ or $\Tgl$ become $0$ as expected when noise was sufficiently small. The SVM work above showed via simulation that as observation window size was increased, the short-term, assumption-less measurements converged to the long-term results. Similar application of $\Ipw$ and $\Tgl$ presents comparable results here.

\section{Results}
\label{sec:results}

\begin{figure}
\centering
  \includegraphics[width=\columnwidth]{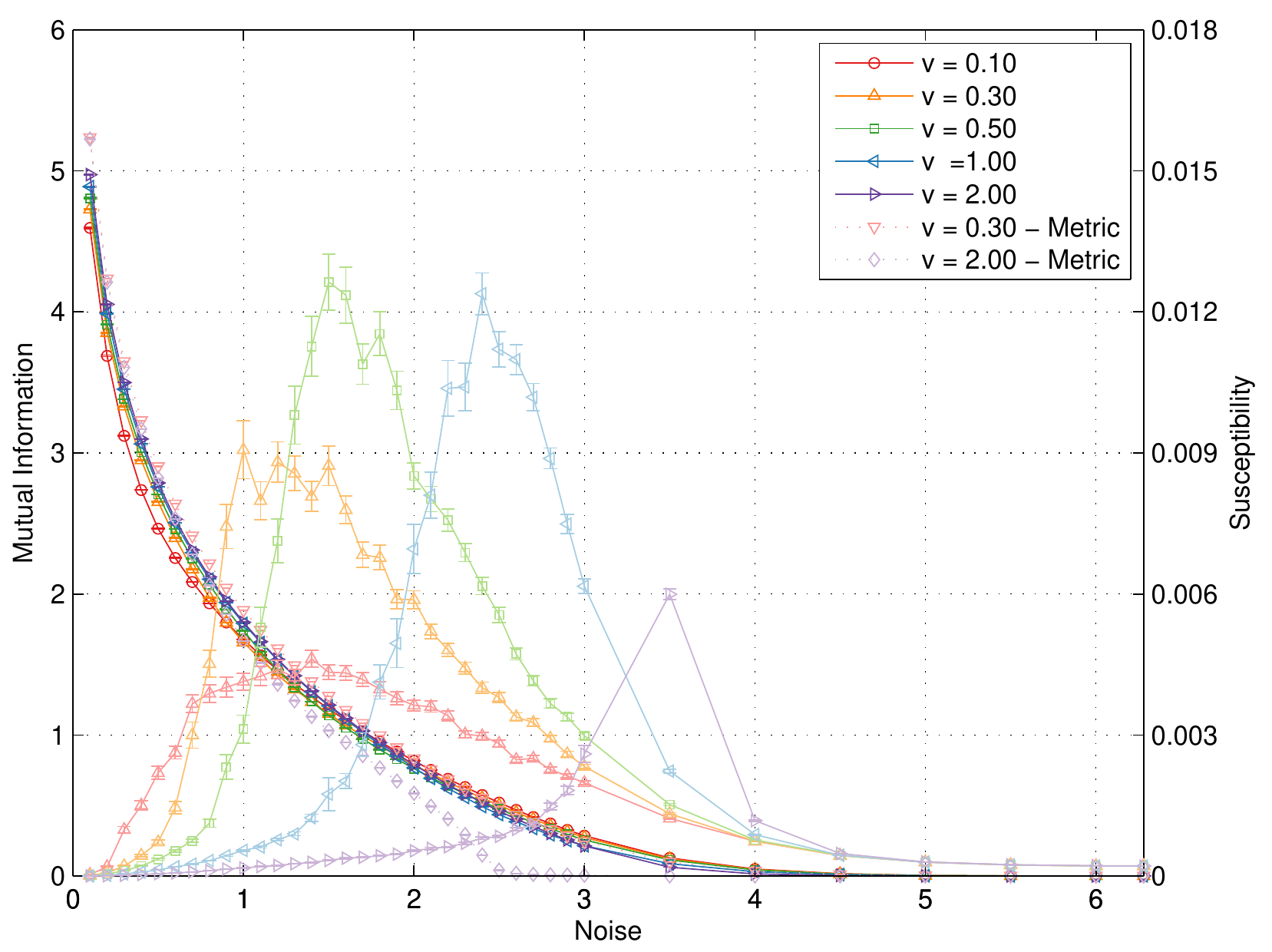}
  \caption{Long-term MI $\IpwLT$ calculated according to \eqnRef{mi1D} for a range of particle velocities. System size $N=1000$ particles with $\rho=0.25$. Statistics constructed from 20 realisations at observation time $T=500$ time steps. Error bars at 1 standard error were calculated from 10 repetitions of the experiment. $\entrop(\Theta_I-\Theta_J)$ was calculated using a 512-bin histogram estimator. Faint lines show susceptibility, $\chi$, of runs.}
  \label{fig:mi1DvelScaling}
\end{figure}

Figure~\ref{fig:mi1DvelScaling} shows the long-term MI, $\IpwLT$, estimated in simulation according to \eqnRef{mi1D}, which behaves similarly to the metric case, particularly, the absence of any peak at the phase transition, and divergence to $\infty$ as $\eta\to 0$.

\begin{figure}
\centering
  \includegraphics[width=\columnwidth]{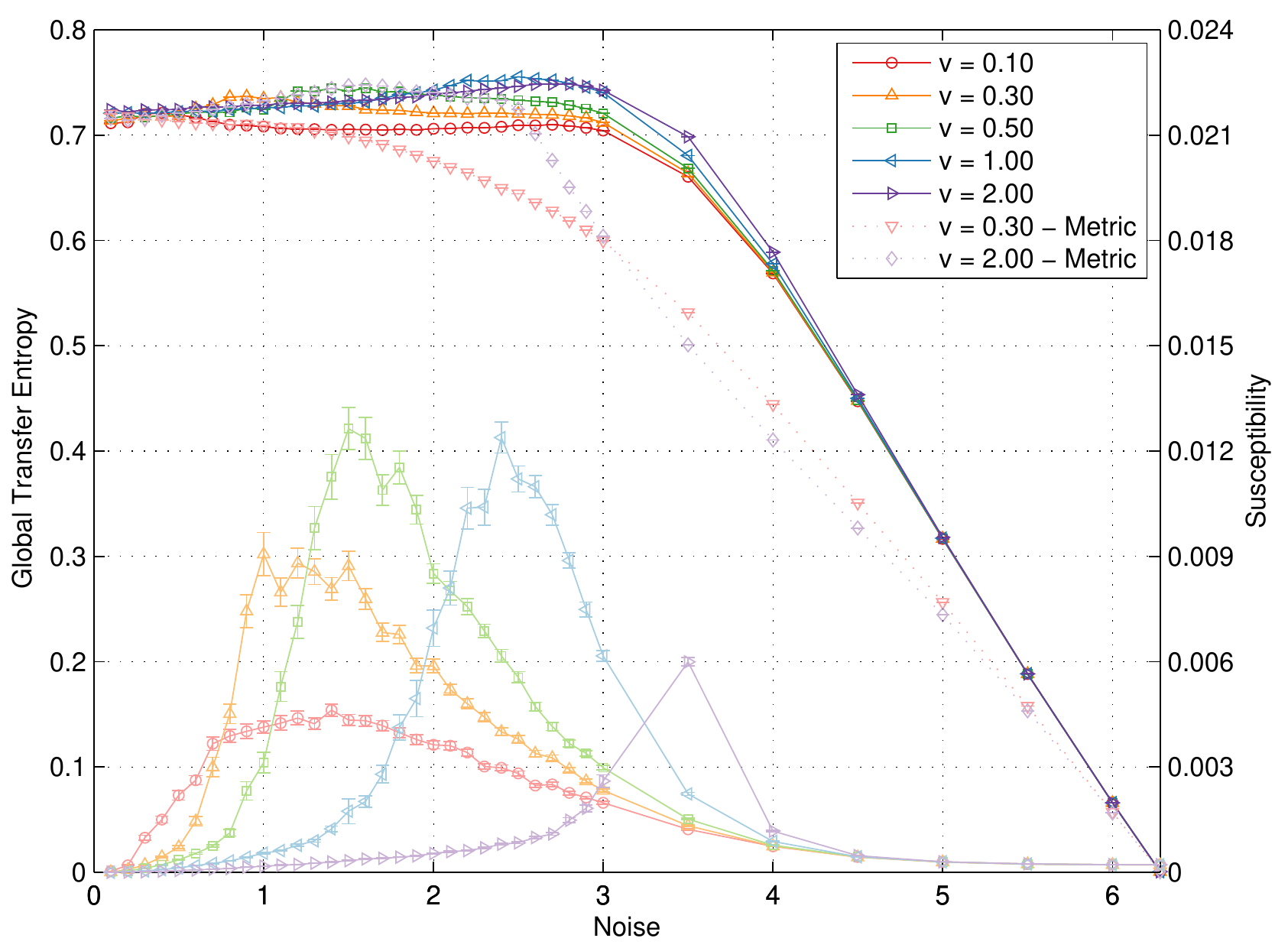}
  \caption{Long-term GTE $\TglLT$ calculated according to \eqnRef{gte1D} for a range of particle velocities using the same data set as \figRef{mi1DvelScaling}. System size $N=1000$ particles with $\rho=0.25$. Statistics constructed from 20 realisations at observation time $T=500$ time steps. Error bars at 1 standard error were calculated from 10 repetitions of the experiment. $\entrop(\Theta'_I-\Theta_I)$ was calculated using a 512-bin histogram estimator. Faint lines show susceptibility, $\chi$, of runs, repeated from \figRef{mi1DvelScaling}.}
  \label{fig:gte1DvelScaling}
\end{figure}

The long-term GTE, $\TglLT$, on the other hand behaves completely differently to the metric case around the phase transition. In the metric case, the noise value, $\eta$, at which $\TglLT$ reached the convergence value scaled with both the velocity and the peak in $\chi$, where higher velocities converged faster---after slightly overshooting around the phase transition for very high velocities ($v=1,2$). Figure~\ref{fig:gte1DvelScaling} shows that in the topological case however, $\TglLT$ reaches the approximate convergence value for \emph{all} velocities around $\eta=3$, regardless of susceptibility peak location, although small peaks do appear to approximately track the phase transition locations. Thus there is a baseline for information flow in the system, dependent on noise, and independent of the phase---the low velocity regime has $\nicesim0.72$ bits of information flow well above the phase transition at $\eta\approx1.5$---with minor peaks in information flow whose locations diverge from the transition location as velocity is taken to extreme values---\ie, $v=0.1, 2.0$.

\begin{figure}
\centering
  \includegraphics[width=0.92\columnwidth]{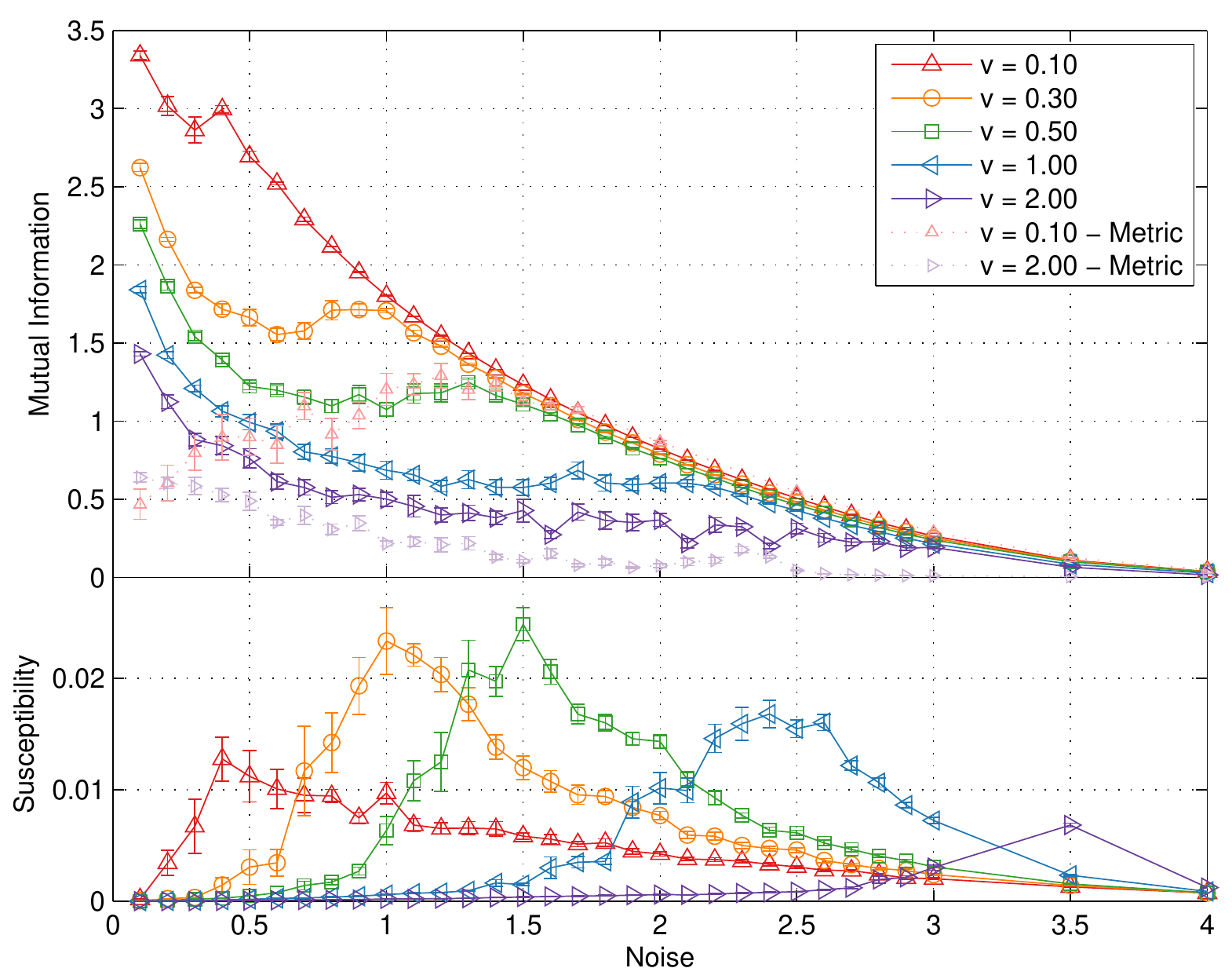}
  \caption{$\Ipw$ (top plot) estimated according to \eqnRef{Ipw} and susceptibility, $\chi$, (bottom plot) for a range of velocities for system size $N=1000$ particles with density $\rho=0.25$. Estimation was performed over $T=5000$ time steps after a relaxation period, using a nearest-neighbour estimator~\cite{Kraskov04}. Error bars at 1 standard error calculated by 10 repetitions. Dotted lines (small symbols) in top plot show previous metric results.}
  \label{fig:miNNvelScaling}
\end{figure}

Estimates for the short term MI and GTE can be seen in \figRefs{miNNvelScaling}{gteNNvelScaling}, respectively, measured with no ergodic assumption. In both cases, local peaks exist at the approximate phase transition---as indicated by a peak in susceptibility---for all velocity values, except $v=2$ for $\Ipw$. Both $\Ipw$ and $\Tgl$ diverge from their long term counterparts below $\eta_c$, with higher velocities diverging further. However at sufficiently low noise, all results converge to the long term results---$+\infty$ for $\Ipw$ and $\nicesim0.7$ bits for $\Tgl$.

\begin{figure}
\centering
  \includegraphics[width=0.92\columnwidth]{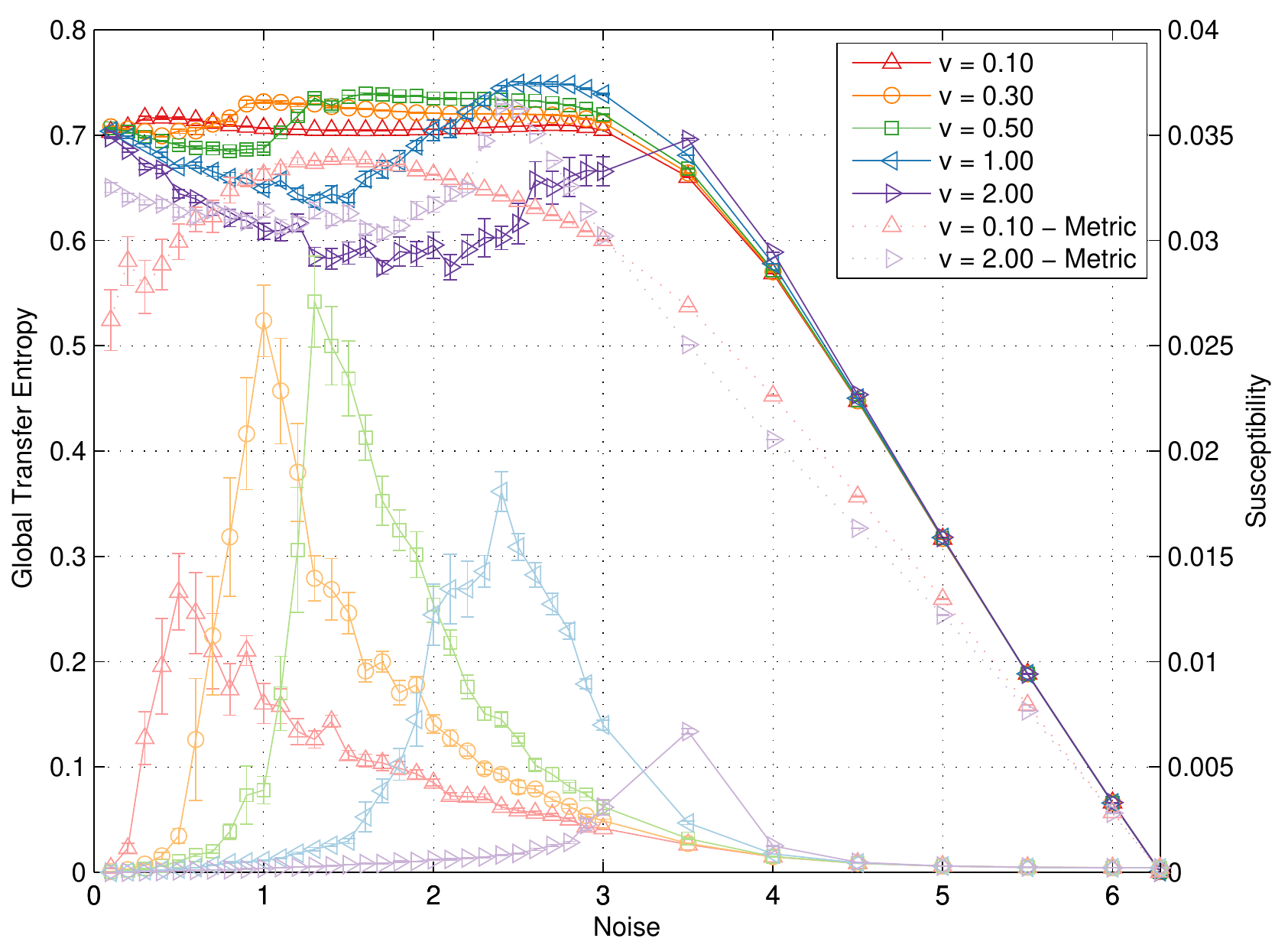}
  \caption{$\Tgl$ estimated according to \eqnRef{Tgl2D} for a range of velocities for system size $N=1000$ particles with density $\rho=0.25$. Estimation was performed over $T=5000$ time steps after a relaxation period, using a nearest-neighbour estimator~\cite{Kraskov04,gomez-herrero15}. Error bars at 1 standard error calculated by 10 repetitions. Dotted lines show previous metric results.}
  \label{fig:gteNNvelScaling}
\end{figure}

\begin{figure}
  \centering
  \includegraphics[width=\columnwidth]{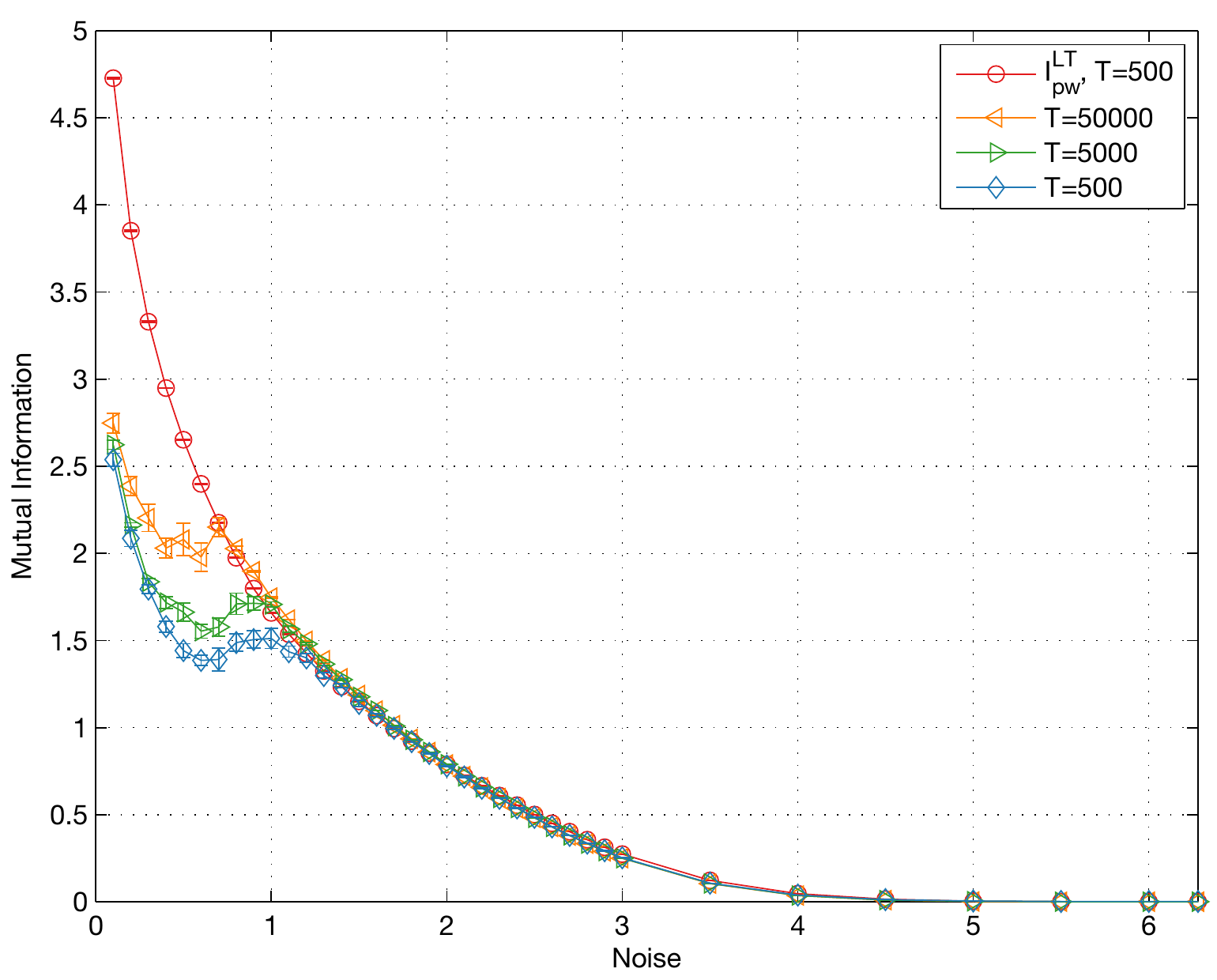}
  \caption{$\Ipw$ estimates according to \eqnRef{Ipw} at fixed velocity $v=0.30$ for a range of observation times, T, as indicated, along with the long-term $\IpwLT$ as per \figRef{mi1DvelScaling}. Other simulations as per previous figures.}
  \label{fig:miNNwindowSize}
\end{figure}

\begin{figure}
\centering
  \includegraphics[width=\columnwidth]{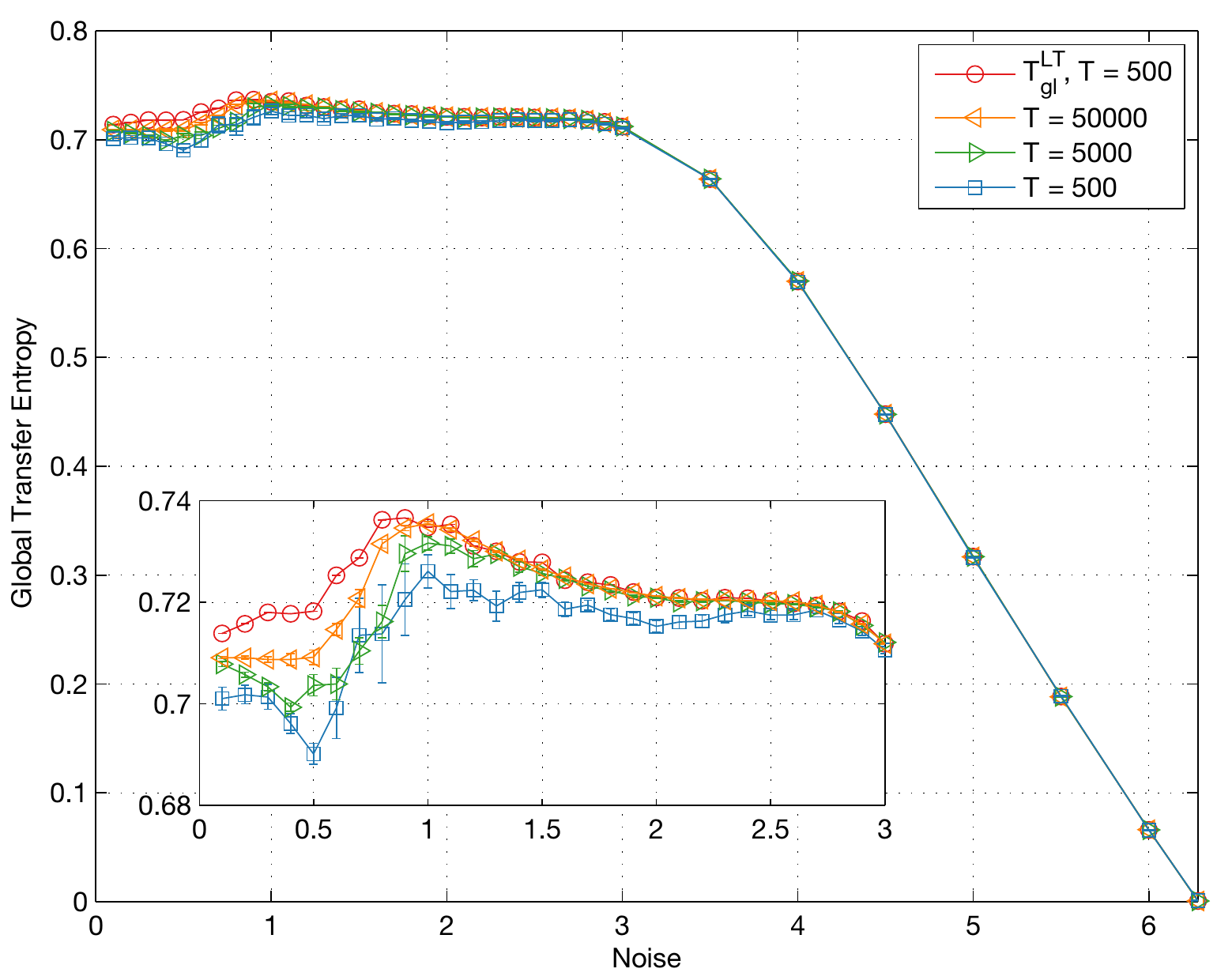}
  \caption{$\Tgl$ estimates according to \eqnRef{Tgl2D} at fixed velocity $v=0.30$ for a range of observation times, T, as indicated, along with the long-term $\TglLT$ as per \figRef{gte1DvelScaling}. Other simulations as per previous figures. Inset: Enlarged plot for $0\leq\eta\leq3$.}
  \label{fig:gteNNwindowSize}
\end{figure}

In contrast, the metric case~\cite{Barnett18,gte-paper} exhibited only minor convergence at low noise (\ie, $\Ipw$ and $\Tgl$ were non-zero but also not approaching the long-term results). Increasing the observation window size of the metric study revealed that this behaviour was caused by the short observation time, where the system was not given enough time to explore the entire phase space available. Figures~\ref{fig:miNNwindowSize}~and~\ref{fig:gteNNwindowSize} repeat this experiment here, showing $\Ipw$ and $\Tgl$ as measured over differing observation window sizes. In both cases, increasing the observation time allows for the assumption-free short term estimations to approach the ergodic long-term behaviours. It remains however, that the short-term topological case converges to the long-term behaviour more quickly than the metric case. Here we take a brief detour to explore the mechanism that leads to this faster convergence before presenting final results and discussion.

% \begin{figure}
%   \includegraphics[width=\columnwidth]{topo_img_final/mi_N_scaling.pdf}
%   \caption{$\Ipw$ estimates according to \eqnRef{Ipw} at fixed velocity $v=0.30$  and fixed observation time $T=5000$ for a range of system sizes, $N$, as indicated along with the long-term $\IpwLT$ as per \figRef{mi1DvelScaling}. Inset: Enlarged plot for $0\leq\eta\leq1.2$}
%   \label{fig:miNNnScaling}
% \end{figure}

% \begin{figure}
%   \includegraphics[width=\columnwidth]{topo_img_final/gte_N_scaling.pdf}
%   \caption{$\Tgl$ estimates according to \eqnRef{Tgl2D} at fixed velocity $v=0.30$  and fixed observation time $T=5000$ for a range of system sizes, $N$, as indicated along with the long-term $\TglLT$ as per \figRef{gte1DvelScaling}. Inset: Enlarged plot for $0\leq\eta\leq3$.}
%   \label{fig:gteNNnScaling}
% \end{figure}

\subsection{Fragmentation}

As mentioned above, $\Ipw$ and $\Tgl$ converge to their respective long term behaviours more quickly---that is, using much smaller observation window sizes---than in the metric case. A brief discussion of why this is the case is presented here (See~\cite{Brown17a}~for additional discussion on the following phenomenon).

\begin{figure}
%onefigure{images/epl_topo_fragmentation_transition.png}
\centering
  \includegraphics[width=\columnwidth]{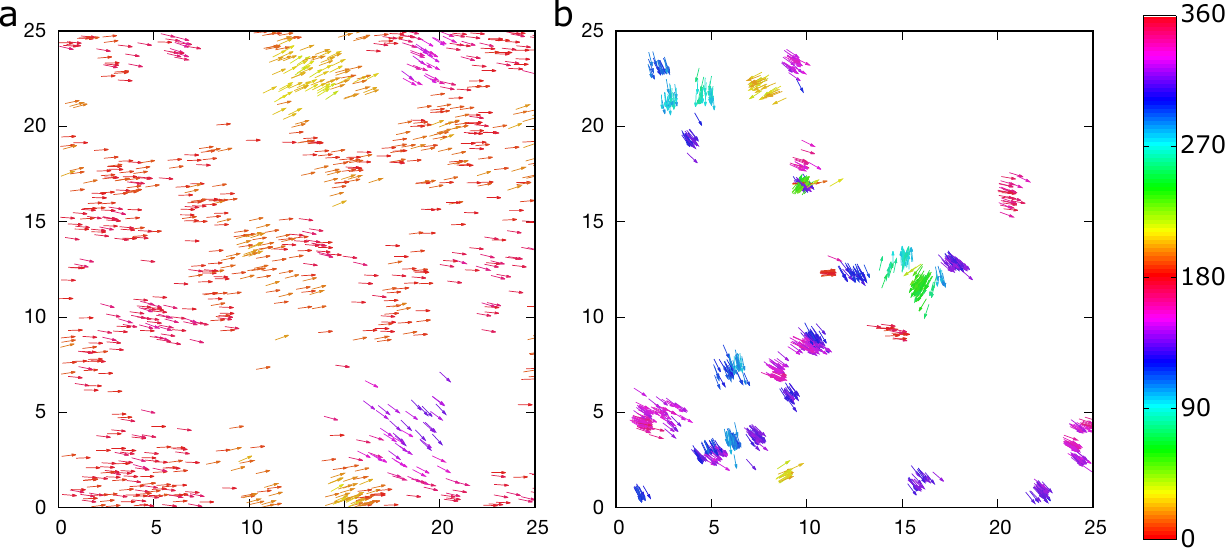}
  \caption{Snapshots of one flock at a) $t=3000~(\varphi=0.97)$ and b) $t=7000~(\varphi=0.75)$. System initialised facing right ($\theta_i(0)=0~\forall~i\in N$), with parameters $N=1000, \rho=1.6, s=0.05, \eta=0.1\pi, k_T=7$ as per \cite{Komareji13}, where a) matches Fig. 1 of \cite{Komareji13}---noting their initial state, $\theta_i(0)=\frac{\pi}{2}$. Particle colour indicates heading angle, such that $\theta=\theta+\pi$. Reprinted from \cite{Brown17a}.}
  \label{fig:consensusSnapshots}
\end{figure}

First, note that the difference in information theoretic quantities between the metric and topological cases occurs mostly at low noise. At and above the phase transition, both cases behave similarly with respect to their long term behaviours. Thus the following discussion will focus on the low noise regime. Visualisation of the topological system at low noise reveals a disordered swarm of small, ordered flocks as seen in \figRef{consensusSnapshots}b.

Komareji and Bouffanais provide a framework for analysing consensus and its resilience in topological systems in \cite{Komareji13} by constructing directed connectivity graphs known as \emph{swarm signalling networks}~(SSN). They demonstrate that for a system to achieve consensus the SSN needs to be a strongly connected super set during most time steps---that is, for the overwhelming majority of time steps, there should exist a path between every pair of flock members in the SSN--- and that to achieve this, $k$ should be $6$ or more, in agreement with bird flocking studies~\cite{Ballerini08}. They also show the case of $k=3$, exhibiting similar behaviours as observed in \figRef{consensusSnapshots}b.

However, while Komareji and Bouffanais \cite{Komareji13} do indeed show consensus for topological systems with $k=6$ neighbours, it is for short time scales, $T=3000$. This consensus is not permanent and over time devolves into the aforementioned quasi-disordered state~\cite{Brown17a}. This work is reprinted here with a system of $N=1000$ particles with density $\rho=1.6$---as used in \cite{Komareji13}---in Fig.~\ref{fig:consensusSnapshots}a. Fig.~\ref{fig:consensusSnapshots}b shows the same system at a future time step, where it has devolved to the quasi-disordered state. Similar behaviours are seen for all $k$ tested, $k \in \{6, 7, 8, 9, 20, 40\}$, as well as $\rho=0.25$ as used throughout the rest of this work. The size of the fragmented flocks scales with $k_T$, while the time taken to devolve scales with $k, s$ and $\rho$.

\begin{figure}
\centering
% \onefigure{images/epl_topo_fragmentation_transition.png}
  \includegraphics[width=0.92\columnwidth]{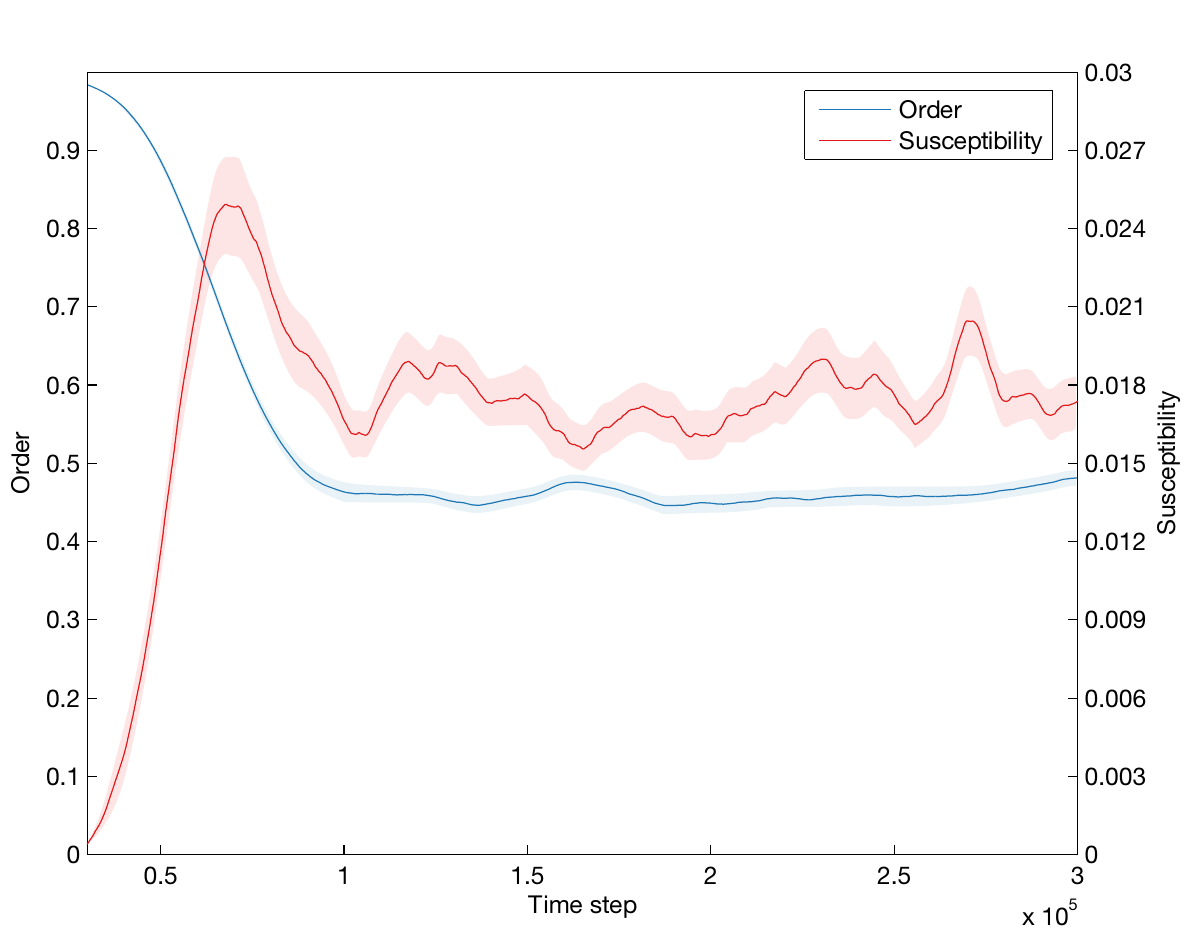}
%Fix box of this image
%\onefigure{epl-template.eps}
  \caption{Plot of order (left axis) and susceptibility, $\chi$, (right axis)---measured over sliding window of $3\e4$ time steps. All particles initialised facing right ($\theta_i(0)=0~\forall~i\in N$), using parameters $N=1000, \rho=0.25, s=0.05, \eta=0.2, k_T=7$. $\rho$ and $\eta$ reduced to increase time taken for fragmentation to occur to better visualise $\chi$. Peak in $\chi$ at $t=7\e4$ indicates flocks fragmenting into a quasi-disordered state as described in text. Non-zero $\chi$ after the peak demonstrates that the quasi-disordered state is maintained for the remainder of the simulation. Measurements repeated 100 times to calculate standard error (shaded regions). Reprinted from \cite{Brown17a}.}
  \label{fig:topoFragTransition}
\end{figure}

Fig.~\ref{fig:topoFragTransition} shows $\varphi$ and $\chi$ of one set of simulations at low $\eta$, for a large time frame. The peak in $\chi$ and behaviour of $\varphi$ indicates the system does indeed experience a phase transition from an ordered state to a disordered one while $\eta$ remains constant. 

The mechanics of this fragmentation arise due to using topological interactions and the Vicsek model relying solely on an alignment rule, with no rule governing particle proximity---for example, some form of magnetic repulsion. Due to the lack of dispersion, it is inevitable for groups of particles to tightly contract around some local centre of mass. Combined with topological interactions, these groups disconnect themselves from the overall SSN reducing it to a weakly connected super set, before a positive feedback loop eventually dissolves the particle flock into compact, disjoint sets~\cite{Brown17a}. 

Interestingly, however, this has the side effect of allowing the flock to more rapidly explore a larger volume of the phase space. Information theoretic quantities are measured on the headings of interacting particles, such that $(X,Y,X') = (\theta_i(t), \theta_j(t), \theta_i(t+1))$. In the low noise metric case, all particles align with the flock direction, resulting in data points such that $X \approx Y \approx X'$ for some heading $0 \leq X \leq 2\pi$. Over time due to thermal fluctuations, the flock heading will proceed on a random walk about the unit circle exploring the complete volume of phase space, which leads to a data set uniform over the $X=Y=X'$ diagonal, as exploited by the the long-term limit approach above.

In the topological case, realisations of $X, Y,$ and $X'$ are still generated from interacting particles, which will still lead predominantly to the $X\approx Y \approx X'$ case since wildly diverging sub-flocks are compact and rarely interact. The key difference however is that the headings of individual sub-flocks are arbitrary and thus each sub-flock generates data along the $X=Y=X'$ diagonal. Furthermore, any collisions between sub-flocks produce large jumps to new areas of phase space (\ie, a novel heading produced from the headings of both flocks). The topological flock is thus able to obtain a more representative exploration of the phase volume more quickly than the metric flock, leading to faster convergence with the long-term limit behaviours.

\begin{figure}
\centering
  \includegraphics[width=0.92\columnwidth]{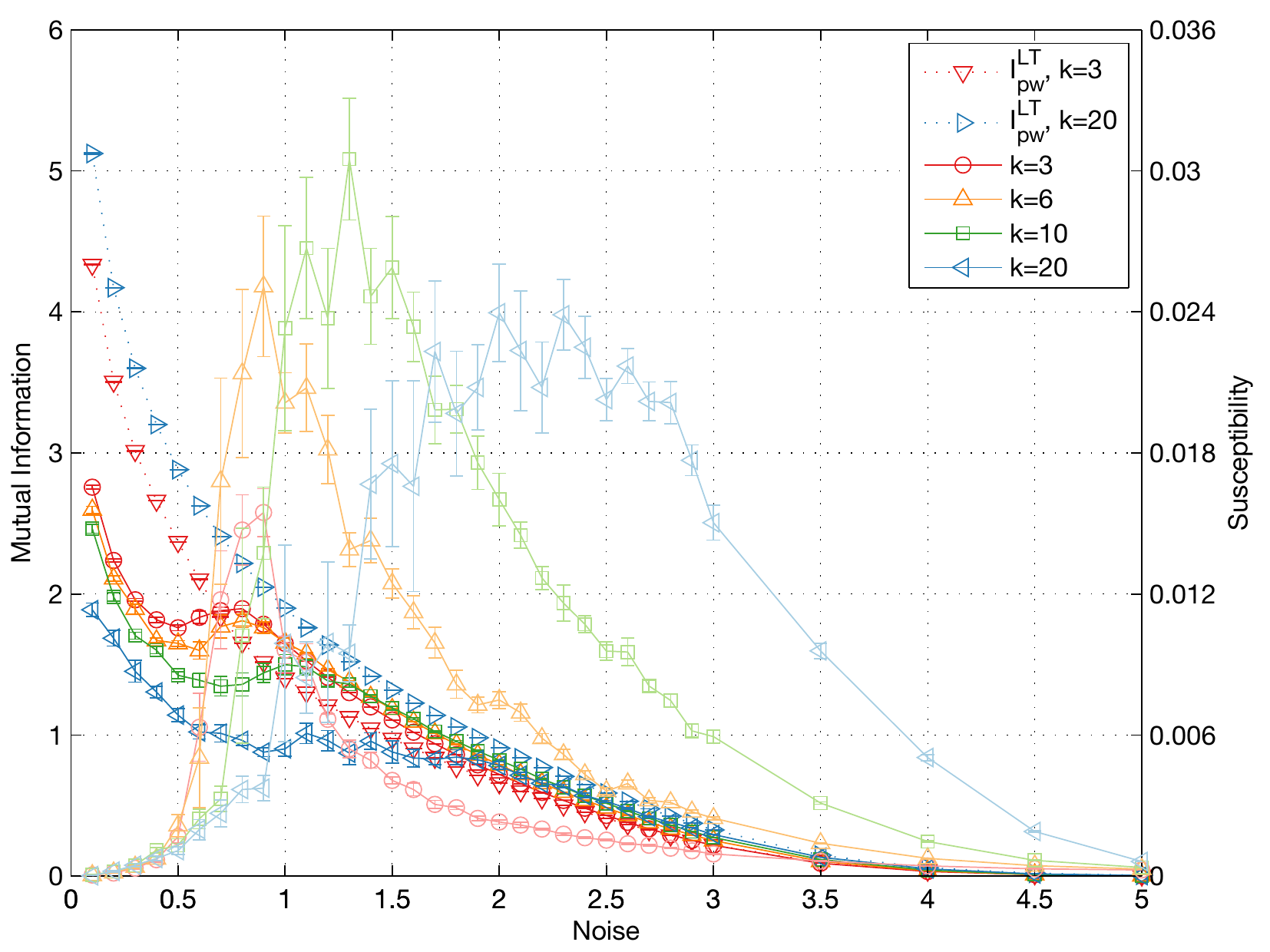}
  \caption{$\Ipw$ estimates according to \eqnRef{Ipw} at fixed velocity $v=0.30$  and observation time $T=5000$ for a range of neighbours, $k$, as indicated along with the long-term $\IpwLT$ as per \figRef{mi1DvelScaling}.}
  \label{fig:miNNkScalingLow}
\end{figure}

\begin{figure}
\centering
  \includegraphics[width=0.92\columnwidth]{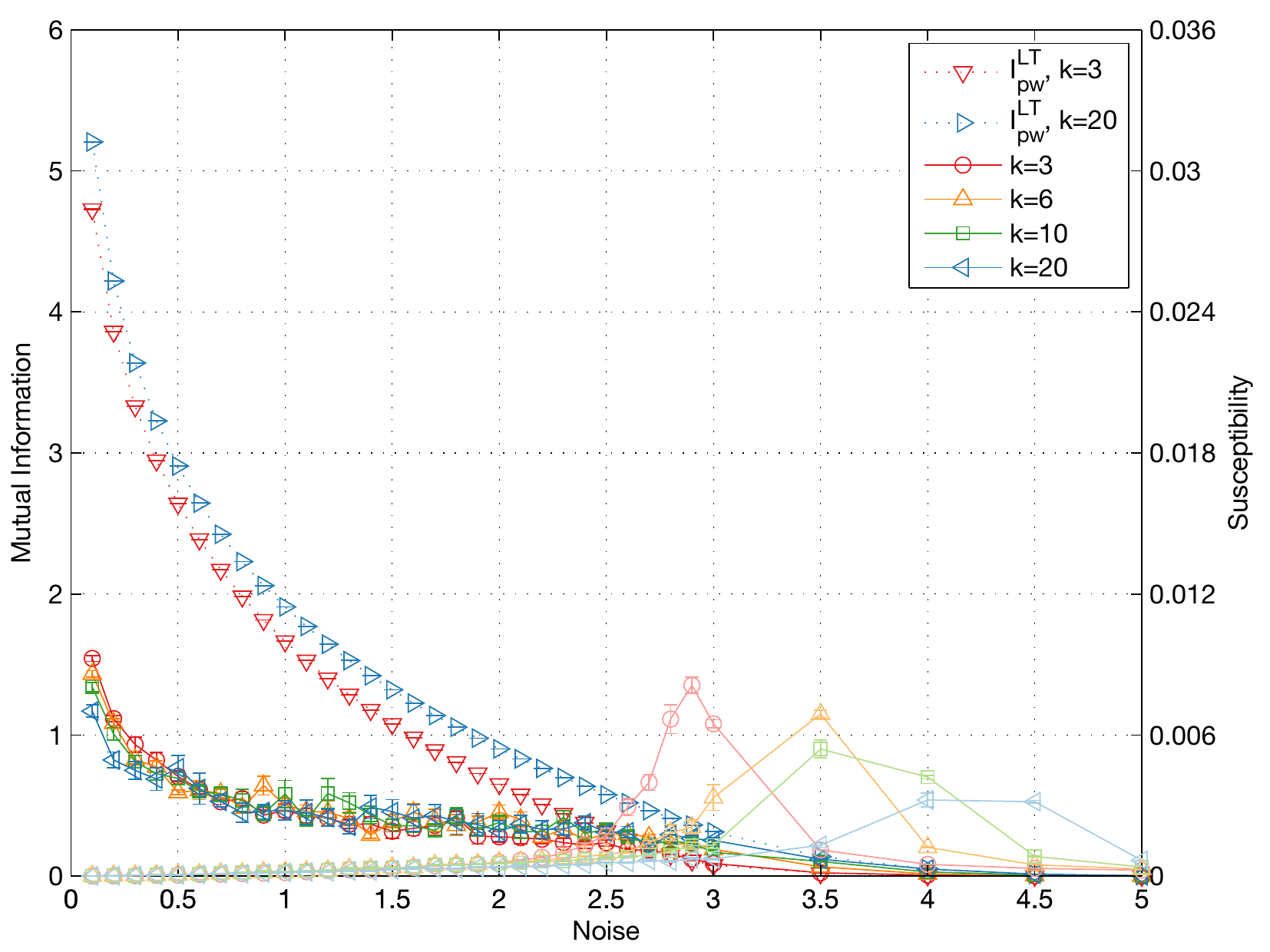}
  \caption{$\Ipw$ estimates according to \eqnRef{Ipw} at fixed velocity $v=2.00$  and observation time $T=5000$ for a range of neighbours, $k$, as indicated along with the long-term $\IpwLT$ as per \figRef{mi1DvelScaling}.}
  \label{fig:miNNkScalingHigh}
\end{figure}

Finally, we return to our remaining results, exploring the effect of varying the number of topological neighbours, $k \in \{3, 6, 10, 20\}$. Figures \ref{fig:miNNkScalingLow} and \ref{fig:miNNkScalingHigh} show short term MI estimated at low and high velocities ($v=0.30, 2.00$, respectively). At low velocity, short term MI does diverge at low noise, approaching the long term limit, with smaller $k$ values tracking the long term limit more closely. This is in agreement with the above fragmentation discussion, where smaller $k$ values lead to more rapid fragmentation, which in turn leads to more efficient exploration of phase space. Conversely, at high velocity, all $k$ values have difficulty approaching the long term limit. This echoes the behaviour in the metric model (See \figRef{miNNvelScaling}). Here, high velocity increases the chance for sub-flocks to ``jump'' past each other, significantly reducing the number of collisions and thus novel headings generated from those collisions.

\begin{figure}
\centering
  \includegraphics[width=0.92\columnwidth]{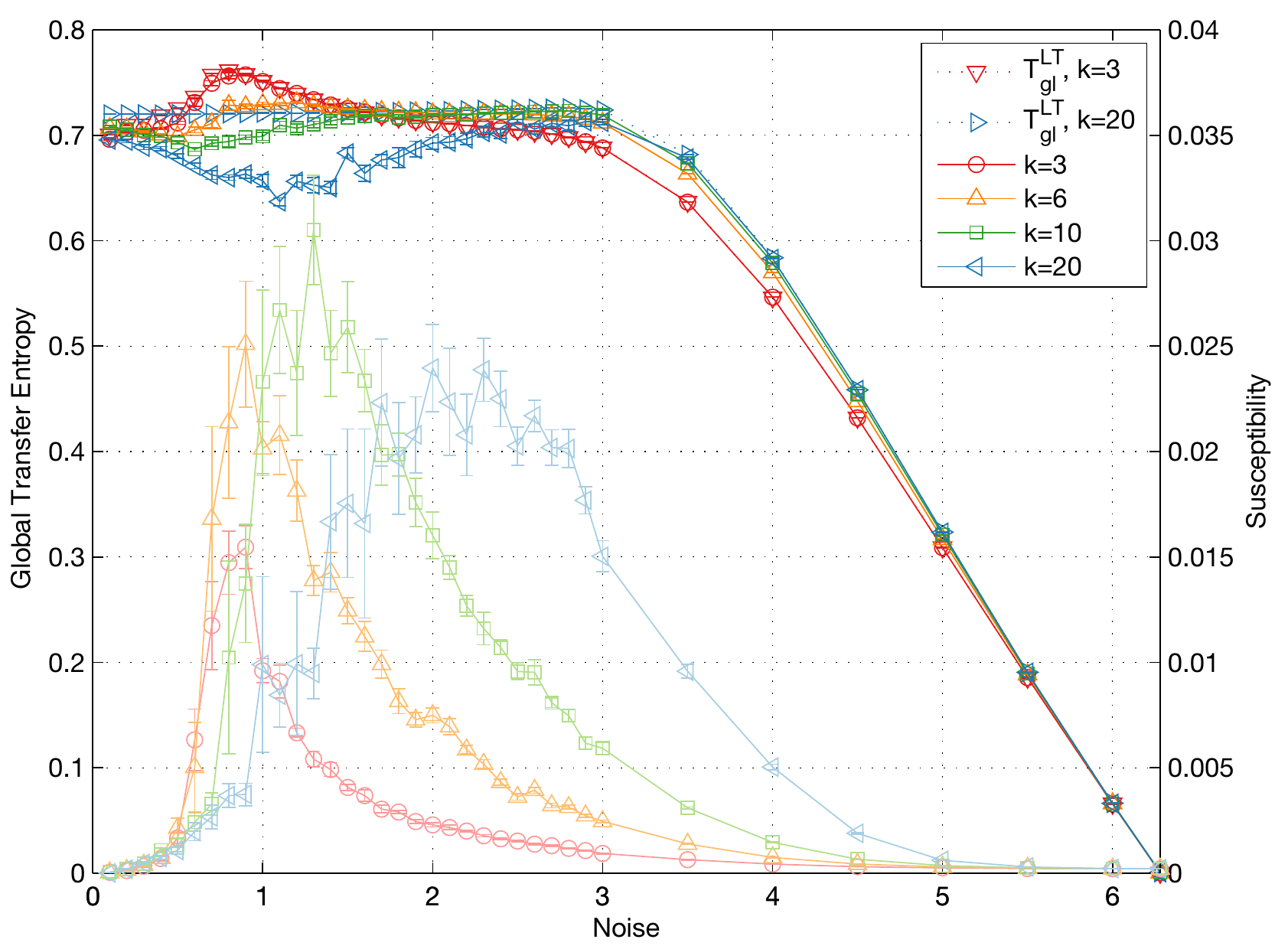}
  \caption{$\Tgl$ estimates according to \eqnRef{Tgl} at fixed velocity $v=0.30$  and observation time $T=5000$ for a range of neighbours, $k$, as indicated along with the long-term $\IpwLT$ as per \figRef{gte1DvelScaling}.}
  \label{fig:gteNNkScalingLow}
\end{figure}

\begin{figure}
\centering
  \includegraphics[width=0.92\columnwidth]{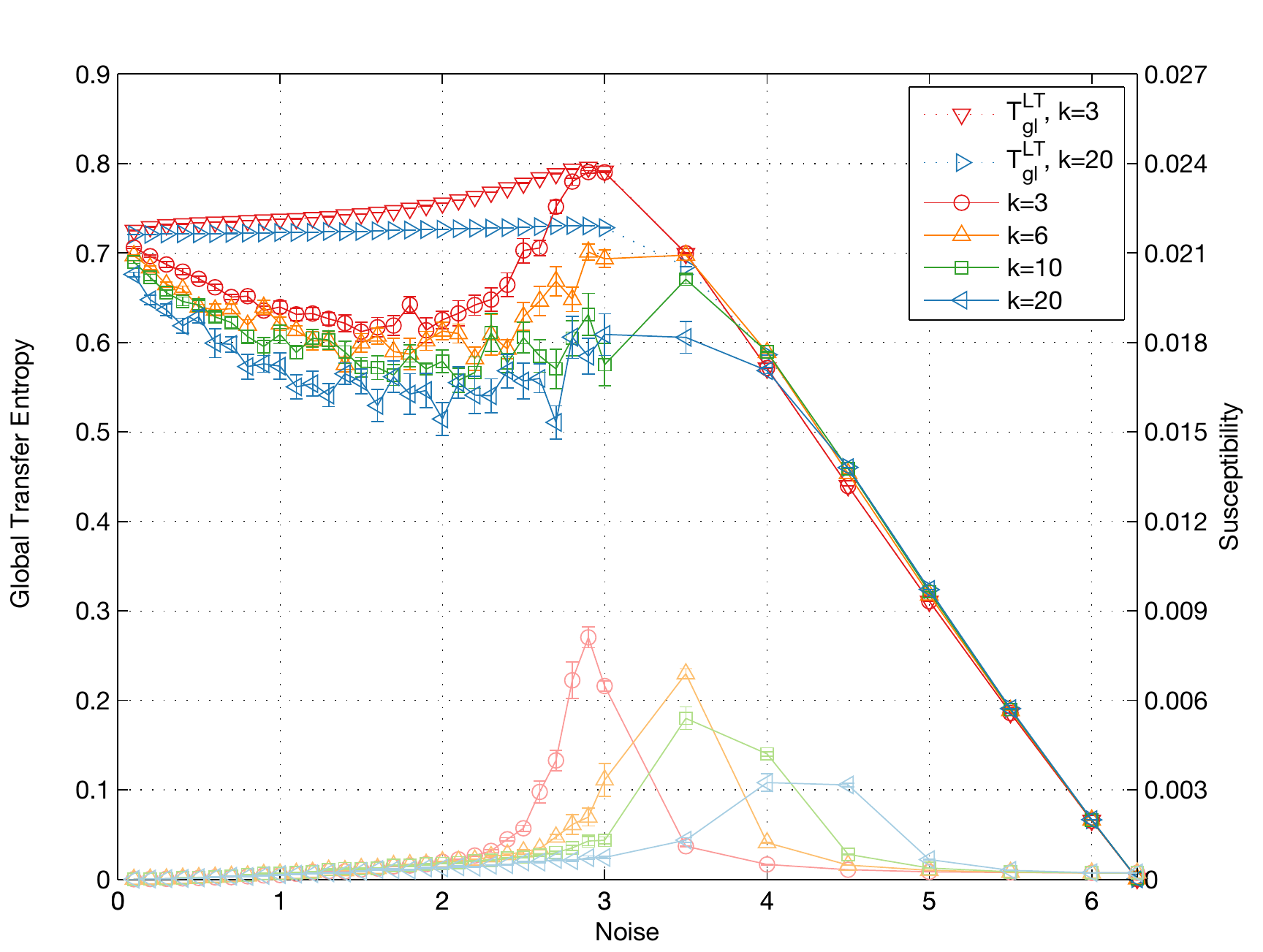}
  \caption{$\Tgl$ estimates according to \eqnRef{Tgl} at fixed velocity $v=2.00$  and observation time $T=5000$ for a range of neighbours, $k$, as indicated along with the long-term $\IpwLT$ as per \figRef{gte1DvelScaling}.}
  \label{fig:gteNNkScalingHigh}
\end{figure}

This pattern of increased convergence towards the long-term limit results appears for the GTE as well. At low velocity (\figRef{gteNNkScalingLow}), short term $k=3$ provides an impressive approximation of the long-term limit. As above, as $k$ increases---and thus fragmentation decreases---convergence to long-term behaviour diminishes. The effect of reduced collisions and novel headings is not as prominent here, as information flow is dependent only on the \emph{difference} in headings in subsequent time steps (\eqnRef{gte1D}), and thus even at high velocity, convergence with the long-term limit is seen (\figRef{gteNNkScalingHigh}).

Of particular note is that as $k$ increases, the long-term GTE becomes flatter for both velocities tested, for all $\eta<\pi$, regardless of where the peak in susceptibility occurs. At these noise levels, a (solitary) sub-flock can never spontaneously flip direction (\ie, $|\phi_i(t) - \phi_i(t+1)| < \frac\pi2~\forall~t$), which maintains the Gaussian distribution of headings about the consensus heading (See \cite{gte-paper} for extended details) giving $\nicesim 0.75$ bits. When other sub-flocks are introduced, spontaneous flips can occur due to collisions, leading to a flurry of information flow, which explains the flattening as $k$ increases: for fixed $N$, increased $k$ produces fewer sub-flocks and thus fewer collisions and flips.

\section{Conclusion} 
\label{sec:conclusion}

Analysis of the topological Vicsek model over short observation windows reveals that information flow more rapidly approaches the long term-limit behaviour in the TVM than its metric counterpart. This phenomenon arises due to an inherent instability in the topological Vicsek model which enables much faster exploration of the total phase volume. Future work would measure the information theoretic quantities of more stable models, perhaps even those measured from real-world flocks.

While the long-term observation window behaviour was similar for MI in both interaction methods, diverging to $+\infty$ as noise decreased, the long-term behaviour of GTE was vastly different. At very low noise, the system still converges to $~0.72$ bits, however, the TVM reaches this value at $\eta\approx\pi$, regardless of the location of the phase transition. Information flow is thus maximal not just at the phase transition and the entire ordered regime as in the SVM, but also for portions of the disordered regime as well, dependent on velocity.

\section{Acknowledgements}
The National Computing Infrastructure (NCI) facility provided computing time for the simulations under project e004, with part funding under Australian Research Council Linkage Infrastructure grant LE140100002.

Joshua Brown would like to acknowledge the support of his Ph.D. program and this work from the Australian Government Research Training Program Scholarship.

Lionel Barnett's research is supported by the Dr. Mortimer and Theresa Sackler Foundation.

\bibliographystyle{plain}

\end{document}